\documentclass[aps,prl,twocolumn,groupedaddress,amssymb,superscriptaddress]{revtex4}
\usepackage[dvips]{graphicx}
\DeclareGraphicsExtensions{.eps,.ps}
\usepackage{bm}      
\newcommand{\ybcox}{Y\-Ba$_2$\-Cu$_3$\-O$_{6+x}$}
\newcommand{\calabalacuo}{Ca$_{1-x}$La$_{x}$Ba$_{2-y}$La$_{y}$Cu$_3$O$_{6+z}$}
\newcommand{\lscozn}{La$_{2-x}$Sr$_x$\-Cu$_{1-y}$Zn$_y$O$_4$}
\newcommand{\ybcoca}{Y$_{1-y}$Ca$_y$\-Ba$_2$\-Cu$_3$\-O$_{6+x}$}
\newcommand{\lsco}{La$_{2-x}$Sr$_x$\-CuO$_4$}
\begin{document}
\bibliographystyle{apsrev}


\title{Nanoscopic coexistence of magnetism and superconductivity in YBa$_2$Cu$_3$O$_{6+x}$ detected by $\mu$SR.}


\author{S.~Sanna} 
\affiliation{Dipartimento di Fisica e Unit\`a INFM di Cagliari, I 09042 Monserrato (Ca), Italy}
\author{G.~Allodi} 

\affiliation{Dipartimento di Fisica e Unit\`a INFM di Parma, I 43100 Parma, Italy}
\author{G.~Concas} 
\affiliation{Dipartimento di Fisica e Unit\`a INFM di Cagliari, I 09042 Monserrato (Ca), Italy}

\author{A.~D.~Hillier}
\affiliation{ISIS Muon Facility, Rutherford Appleton Laboratory, Chilton, OX11 0QX,  United Kingdom}

\author{R.~De Renzi} 
\affiliation{Dipartimento di Fisica e Unit\`a INFM di Parma, I 43100 Parma, Italy}

\date{\today}

\begin{abstract}
We performed zero and transverse field  $\mu$SR experiments on a large number of Y\-Ba$_2$\-Cu$_3$\-O$_{6+x}$ samples. We detect coexistence of antiferro-type (AF) short range magnetism with superconductivity  below $T_f\alt 10$ K in compositions  $0.37\alt x \alt 0.39$. Most muons experience local AF fields, even when SQUID detects a full superconducting volume fraction, which points to a local {\em minimal interference} organization of short AF stripes embedded in the superconductor. A detailed phase diagram is produced and the consequences of the minimal interference are discussed. 
\end{abstract}
\pacs{74.25.Ha;74.62.Dh;74.72.Bk;76.75.+i}
\keywords{Magnetism; High T_c superconductors; Muon spin rotation}

\maketitle




The appearance of low temperature magnetism in low doping cuprate superconductors was an early $\mu$SR claim \cite{brewer:1988} for \ybcox, later also found \cite{Weidinger:1989} in \lsco, but initially highly disputed\cite{Harshman:1989,Heffner:1989}. More recently it has been established that superconducting \ybcoca\  \cite{Niedermayer:1998}, \calabalacuo\  \cite{Kanigel:2002,Keren:2003}, \lscozn\  \cite{Panagopoulos:2002} and \lsco\  \cite{Savici:2002} exhibit magnetic order. Zero field $\mu$SR detects internal spontaneous magnetic fields $\bm{B}_i$ in sizeable volume fractions of samples which display bulk superconductivity (SC). In a few cases the whole volume is involved, i.e. all implanted muons experience internal fields.

Neutron scattering also detects magnetic correlations, notably  \cite{Mook:2002} in YBa$_2$Cu$_3$O$_{6.35}$, exclusively dynamic in nature, although their static counterpart could be elusive due to a very short correlation length. The doubled magnetic unit cell indicates an antiferromagnetic (AF) structure, with a suggested stripe-like character. 

In all the samples explored so far by $\mu$SR  \cite{Niedermayer:1998,Kanigel:2002,Keren:2003,Panagopoulos:2002,Savici:2002} it has been hinted that the cluster spin glass nature of magnetism (low spin freezing temperature, $T_f$, large distributions of $\bm{B}_i$ and absence of long range order as from neutron diffraction \cite{Cheong:1991,Yamada:1998}) might be favored by the disorder inherent in cation substituted perovskites, which directly influences the CuO$_2$ layers.  Conversely in YBa$_2$Cu$_3$O$_{6+x}$ (123, hereafter) the source of disorder, namely the basal CuO$_x$ layers, are farther removed from the  CuO$_2$ layers, but systematic $\mu$SR data were lacking, prior to the present work. Our aim is to clarify whether the appearance of coexisting superconducting and magnetic properties is indeed intrinsic to the unperturbed underdoped CuO$_2$ layers and whether the two properties cooperate or interfere.

We performed $\mu$SR measurements on twenty-four polycrystalline 123 samples (Y$_{1-24}$) prepared by the topotactic technique, which consists of oxygen equilibration of stoichiometric quantities of the two end member specimens, tightly packed in sealed vessels \cite{Manca:2001}. Low temperature annealing yields high quality homogeneous samples with an absolute error of $\delta x\!\!=\!\!\pm 0.02$ in oxygen content per formula unit and a much smaller relative error between samples of the same batch. The width of the interval were the resistance drops from 90\% to 10\% of the onset value is 0.5 K at optimal doping and 6-7K at $x\!\!\alt\!\! 0.4$ ({\em vs.~e.g.}~10K in Ref.\onlinecite{Mook:2002}). The hole content $h$ was determined from the resistive $T_c$ for the superconducting samples \cite{Tallon:1995}, and from the Seebeck coefficient $S$ at 290 K for the non-superconducting ones, using the exponential dependence\cite{Tallon:1995} of $S$ on $h$, with fit parameters determined from our series of samples \cite{Sanna:2004}.
Samples Y$_{1-8}$, with oxygen content $0.20\!\!\le\!\! x\!\!\le\!\! 0.32$ and hole content per planar Cu atom $0.033 \!\!\le\!\! h\!\!\le\!\! 0.055$, never superconduct, and their AF properties were reported previously \cite{Sanna:2003}. Samples Y$_{9-24}$, with $0.32\!\!\le\!\! x\!\!\le\!\!0.42$ and $0.055\!\!\le\!\! h\!\!\le\!\! 0.08$, are superconductors and are the subject of the present work. 

The $\mu$SR experiment were performed on the MUSR spectrometer of the ISIS pulsed muon facility, where the external magnetic field $\bm{H}$ may be applied either parallel to the initial muon spin $\bm{S}_\mu$, in longitudinal (LF) or perpendicular to  $\bm{S}_\mu$, in transverse field (TF) experiments\cite{Schenck:1986}. The LF detector setup is used also in zero field (ZF) experiments.
The asymmetry in the muon decay is always obtained from the count rates $N_i$ in two sets of opposite detectors as ${\cal A}(t) \!\!=\!\! [N_1(t)-\alpha N_2(t)]/[N_1(t)+\alpha N_2(t)]$; the effective relative count efficiency $\alpha$ is calibrated at high temperature\cite{Schenck:1986}.

The spin precessions around the local field $\bm{B}_\mu\!\!=\!\!\mu_0(\bm{H}+\bm{M}_{\mu})$ yields a TF asymmetry as ${\cal A}_T (t) \!\!=\!\! a_{TF} G_{xx}(t)\cos(2\pi\gamma B_\mu t) $, where $a_{TF}$ is the amplitude, $\gamma\!\!=\!\!135.5$ MHz/T is the muon gyromagnetic ratio and $G_{xx}$ the transverse muon relaxation function, i.e.~the time decay of the precession. Superconducting and magnetically ordered phases give rise to a different local magnetization $\bm{M}_\mu$, hence to distinctive features in  $G_{xx}$. The amplitudes of these distinctive signals are directly proportional to the volume fraction where the corresponding order parameter is established.

The ZF asymmetry in the paramagnetic phase is ${\cal A}_Z(t) \!\!=\!\! a_{ZF} G_{zz}(t)$, where the weak longitudinal relaxation $G_{zz}$ originates from the nuclear dipolar fields. No specific feature appears below a SC transition. Magnetic order (whereby in ZF the local muon field $\bm{B}_\mu$ coincides with the internal field $\bm{B}_i$) gives rise to 
\begin{equation}
{\cal A}_Z(t) = a_L G_{zz}(t) +  a_T G_{xx}(t)\cos(2\pi\gamma |{\bm{B}_i}| t), 
\label{eq:Zasymmetry}
\end{equation}
with $a_L+a_T\!\!=\!\!a_{ZF}$, distinguishing the longitudinal ($\bm{B}_i\!\!\parallel\!\! \bm{S}_\mu$, amplitude $a_L$) and the transverse  ($\bm{B}_i\perp \bm{S}_\mu$, amplitude $a_T$) components, referred to the internal field $\bm{B}_i$.
A simple geometric argument for the weights $w_{L,T}\!\!=\!\!a_{L,T}/a_{ZF}$ predicts $w_T=2w_L=2/3$ in polycrystalline samples. If only part of the sample is magnetically ordered the weight becomes $w_L>1/3$ and one can evaluate the volume fraction in which muons experience a net (AF) field $\bm{B}_i$ as $f_{AF}\!\!=\!\!3w_T/2\!\!=\!\!3(1-w_L)/2$. In the AF phase of 123 two $\mu$ stopping sites  are detected \cite{Gluckler:1989,Sanna:2003}, with distinct internal fields, but indistinct longitudinal terms. 

 \begin{figure}
\includegraphics[width=0.37\textwidth,angle=90]{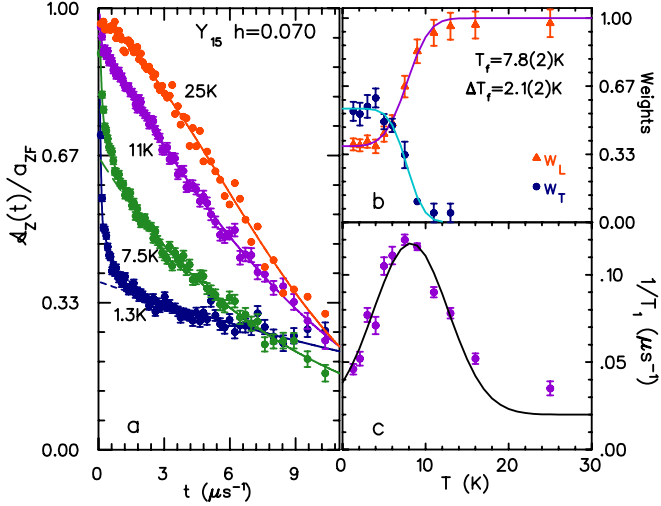}%
 \caption{ZF $\mu$SR, sample Y$_{15}$ a) Normalized asymmetries at $T/T_f\approx 0.17, 0.96, 1.4, 3.2$; b) longitudinal  and transverse  weights $w_L(T), w_T(T)$ with best fit (see text); c) longitudinal relaxation rate, $T_1^{-1}$, vs. temperature; the solid line is the best fit to a simple model \cite{Bloembergen:1948}.}
 \label{fig:1}
 \end{figure}

The combination of ZF and TF experiments on our superconducting samples, Y$_{9-24}$, yields a complete picture of their low temperature electronic properties. 
The presence of {\em static} magnetism in superconducting samples is demonstrated by the appearance of strong magnetic relaxations in the ZF asymmetries, quenched by longitudinal fields larger than a few tens of mT.  A clear example is given by sample Y$_{15}$, with zero resistance at $T_c\!\!=\!\!30(1)$ K and $h\!\!=\!\!0.070(1)$, whose normalized asymmetry ${\cal A}_Z/a_{ZF}$ at $T= 1.3$ K (Fig.~\ref{fig:1}a) displays a very fast Gaussian relaxing transverse component, which is the signature of a distribution of sizeable static internal fields. The best fit yields two Gaussian contributions of standard deviation $\sigma/2\pi\gamma\!\!=\!\!(\overline{B}^2- \overline{B^2})^{1/2}=25$ mT and 5 mT respectively, which correspond roughly to the internal fields at the two $\mu$ sites. For $T\rightarrow$0  their total weight, $w_L$,  is very close to the 1/3 value expected for a fully ordered material (see dashed curves in Fig.~\ref{fig:1}a). This proves the presence for $T<T_f$ of local magnetic fields from static moments {\em throughout the whole sample}, although it does not exclude the simultaneous presence of superconducting carriers, to which ZF $\mu$SR is insensitive.

The magnetic transition is signaled by the appearance of a peak in the longitudinal relaxation rate $T_1^{-1}$ at $T_f$, shown in Fig.~\ref{fig:1}c and obtained from a fit to Eq.~\ref{eq:Zasymmetry} with $G_{zz}\!\!=\!\!exp(-t/T_1)$. It is due to the freezing of spin dynamics upon formation of a cluster spin glass state (at lower dopings two distinct peaks are observed\cite{Sanna:2003}, the second being due to critical slowing down of spin fluctuations at the N\'eel temperature $T_N$).

An independent determination of $T_f$ is obtained from weights. Figure \ref{fig:1}b shows that the transverse weight $w_T(T)$ disappears above $T_f$ (above $T_N$, for lower dopings), whereas the longitudinal weight $w_L(T)$ tends to unity. Their smooth dependence indicates the presence of a distribution of transition temperatures. The solid curves in the figure are a fit to a Gaussian distribution model \cite{Sanna:2003}, yielding $T_f\!\!=\!\!7.8(2)$ K and a width of the distribution of $\Delta T_f\!\!=\!\!2.1(2)$ K. 

\begin{figure}
\includegraphics[width=0.37\textwidth,angle=90]{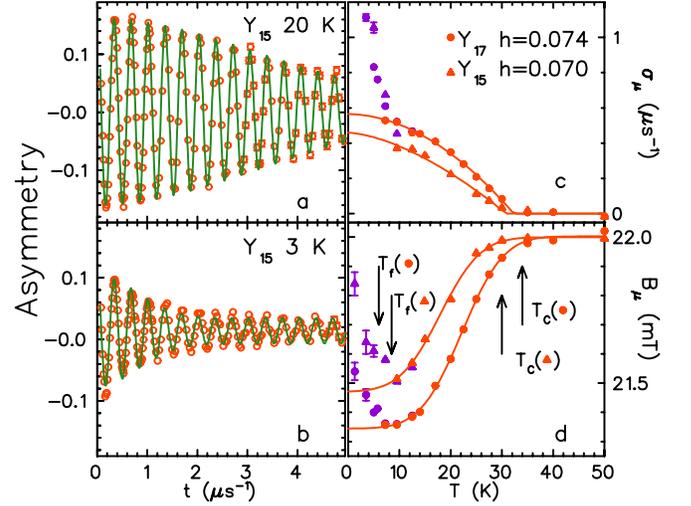}%
 \caption{TF $\mu$SR data ($\mu_0 H\!\!=\!\!22$ mT).  Asymmetry a)  for $T_f\!\!<\!\!T\!\!=\!\!20\mbox{K}\!\!<\!\!T_c$ and b)  for $T\!\!=\!\!3\mbox{K}<T_f$ in sample Y$_{15}$; solid curves are best fits to a Gaussian damped precession. c) Relaxation rate  $\sigma_\mu$ and d) internal field $B_\mu$  from the best fits for samples Y$_{15}$ and Y$_{17}$.}
 \label{fig:2}
 \end{figure}

The TF experiment on the very same sample quantifies the $\mu$SR assessment of superconductivity.
In the normal (paramagnetic) phase the local field $B_\mu$ coincides with $\mu_0 H$ and $G_{xx}(t)$ represents the weak Gaussian relaxation of width $\sigma_n$ due to the random nuclear magnetic dipoles; in the superconducting phase a flux lattice is established  and $G_{xx}(t)\!\!\approx\!\! exp[-(\sigma_\mu^2+\sigma_n^2)t^2/2]$ reflects the corresponding field distribution\cite{Pumpin:1990}, approximated by a sizeable Gaussian decay parameter $\sigma_\mu$. Figure \ref{fig:2}a shows this damped precession.

The temperature dependence of the fitted parameters $\sigma_\mu$ and $B_\mu$ is shown in Fig.~\ref{fig:2}c and d, respectively, together with those of the slightly more highly doped Y$_{17}$ sample ($h$=0.73(1)). They reveal the typical features of powder $\mu$SR data in the SC state, including a clear diamagnetic shift of the local field $B_\mu\!\!=\!\! \mu_0 H(1+\chi)$, with $\chi<0$. The solid lines in Fig~\ref{fig:2}c are best fits to an effective BCS-like temperature power law \cite{Pumpin:1990} for $\sigma_\mu(T), T\!\!>\!\!T_f $, which allows the extrapolation of  the $\sigma_\mu(T\!\!=\!\!0)$ value and the evaluation of the transition temperature $T_{c\mu}$, which coincides with $T_c$ from zero resistance ($R(T)\!\!=\!\!0$) within $\pm 1$ K. The signal amplitude $a_{SC}$ is very close to that of the normal phase ($a_0$, for $T>T_c$), and the volume fraction corresponding to the flux lattice signal is directly obtained as $f_{SC}=a_{SC}/a_0$, with a value remarkably close to unity. 

Spontaneous local fields $\bm{B}_i$ appear for $T<T_f$. Since in polycrystalline samples they are randomly oriented with respect to $\mu_0\bm{H}$, the vector composition $B_\mu\!\!=\!\! |\mu_0\bm{H} + \bm{B}_i|$ leads to a spread of precession frequencies resulting in a much stronger damping of the asymmetry. The experimental data are shown in Fig.~\ref{fig:2}b with the same single Gaussian fit used for $T>T_f$ (solid line). The true lineshape is not Gaussian and a loss of initial asymmetry is also evident. However both $\sigma_\mu(T)$ and $B_\mu(T)$ (Fig.~\ref{fig:2}c,d) clearly reveal the onset of magnetic order. 

 \begin{figure}
\includegraphics[width=0.45\textwidth]{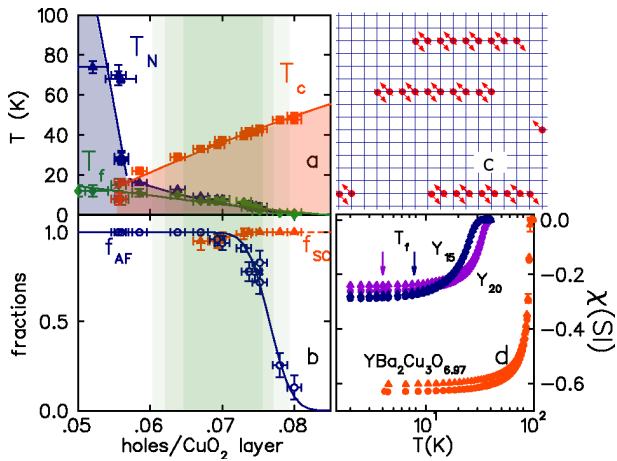} 
 \caption{Samples Y$_{8-24}$. a) magnetic transition temperatures ($\blacktriangle $ from $w_L$, $\blacklozenge$ from $T_1^{-1}$) and SC critical temperature $T_c$, vs. hole concentration $h$; three samples show a distinct $T_N \!\!> \!\!T_f$.   b) Muon volume fractions vs. $h$: AF ($\circ$, at $T\!\!=\!\!0$ K) and SC ($\blacktriangle$, for $T_f\!\!\le\!\! T\!\!\le\!\! T_c$. Lines in a), b) are guides to the eye. c) Sketch of a stripe superconductor: the grid shows the CuO$_2$ plaquettes in the superconducting layer. d) SQUID susceptibility in YBa$_2$Cu$_3$O$_{6.97}$, Y$_{15}$ ($h$=0.70) and Y$_{20}$ ($h$=0.75): $\mu_0 H\!\!=\!\!0.1$ mT, FC ($\blacktriangle$) and ZFC ($\bullet$). }
 \label{fig:3}
 \end{figure}

ZF and TF results impose that {\em nearly  all} implanted muons detect {\em both} a flux lattice for $T_f<T<T_c$ and an internal AF field for $T<T_f$.
Qualitatively similar results are obtained throughout the whole series of samples. Figure \ref{fig:3}a shows the dependence of the magnetic temperature $T_f$ upon hole content $h$ across the coexistence region. A few low doping samples ($h<0.055$) display separate  $T_N$ and $T_f$ (two peaks\cite{Sanna:2003} in $T_1^{-1}$). Also shown is $T_c$ corresponding to the midpoint of the SC transition measured by resistance $R(T)$. The smooth, continuous behavior of all three temperatures demonstrates the well defined electronic properties of our samples and their high reproducibility.

The issue is whether bulk superconductivity survives below $T_f$. It may well, since the coherence length \cite{Sonier:1997,Miller:2003} is $\xi\approx 80-200$ \AA\ for $x=0.60-0.57$, much larger than the characteristic length scale of the SC stripes mentioned above. Although the presence of SC below $T_f$ might be retrieved from $\mu$SR lineshape analysis, as it was attempted by modeling the complex internal field distribution in a single crystal  \cite{Savici:2002}, this is beyond the scope of our polycrystalline data and we reverted directly to SQUID magnetization measurements.

Figure \ref{fig:3}d shows field cooled (FC) and zero field cooled (ZFC) susceptibilities, $\chi\!\!=\!\!M/(H-NM)$, of Y$_{15}$, Y$_{20}$ and of a YBa$_2$Cu$_3$O$_{6.97}$ sample, for reference, in a low field $\mu_0 H\!\!=\!\!0.1$ mT. We assume $N\!\!=\!\!1/3$ for the nearly spherical grain powders of average size $R$$\approx$1 $\mu$m and neglect flux pinning \cite{Nagano:1993}, since $\chi_{ZFC}\!\!\approx$$\chi_{FC}$. With these assumptions the field penetrates a spherical crust of thickness equal to the penetration depth $\lambda$ (extracted from the width, $\sigma_\mu=7.58\cdot10^{-8}\lambda^{-2}$ in SI units\cite{Barford:1988}) and from the ratio of shielded to total volumes we expect $\chi\!\!=\!\!\chi_0(1-f)^3$ with $\chi_0\!=\!-1$ and $f\!=\!\lambda/R$, in good agreement  \cite{Sanna:2004} with the data. No appreciable contribution to $\chi$ is expected from an AF structure in low fields, below $T_f=7.8(2),4(1)$ K, but a change in the superconducting volume would show up in the data, and it does not. Hence the {\em whole volume} remains superconducting down to 2 K, well below $T_f$, i.e. superconductivity is not modified by the onset of static AF correlations. 

The colored stripe in Fig.~\ref{fig:3}a,b shows the region where $T_f\!\!<\!\!T_c$. We evaluate the fraction of muons experiencing AF local fields as $f_{AF}\!\!=\!\!3(1-w_L)/2$ (Fig. \ref{fig:1}b). Likewise the TF amplitude for $T_f\!\!<\!\!T\!\!<\!\!T_c$ yields the fraction $f_{SC}$ of muons stopped in a SC environment. Figure \ref{fig:3}b displays the two fractions, which are both close to unity for seven distinct samples (Y$_{14}$-Y$_{20}$) in a considerable portion of the colored area.

There is no contradiction in these two findings since $f_{AF}\!\!\approx\!\! 1$ does not mean a localized moment at each Cu site. It merely indicates that each muon site experiences non vanishing internal magnetic field, i.e. that all implanted muons are within a short distance $d$ from static magnetic moments. Estimates \cite{Niedermayer:1998,Savici:2002} with dipolar fields ($\propto\!\!mr^{-3}$, $m\!\!\approx\!\!0.6 \mu_B$), yield $d\!\!\approx\!\! 20$ \AA, hence nanoscopic magnetic clusters must be present {\em within} the superconducting volume. Savici {\em et al.}\cite{Savici:2002} show that for circular clusters  a large volume fraction $V_{Cu}>0.8$ of ordered Cu moments would be required to reproduce the observed $100\%$ of magnetic sites, thus resulting in $f_{SC}< 0.2$.  Hence our values $f_{SC}\approx f_{AF}\approx 1$ (Fig.~\ref{fig:3}b) are not compatible with a circular shape; rather they require {\em linear} clusters, where $V_{Cu}$ may be much smaller than $f_{AF}$. An example is shown in Fig~\ref{fig:3}c: fragments of chain separated by superconducting stripes of width $2d$ result in $f_{SC}\!\!\approx\!\! 1-a/d$ ($a=3.8$ \AA\  is the lattice spacing), with all $\mu$  sites within a distance $d$ from Cu moments, within a percolating SC infinite cluster.


\begin{figure}
\includegraphics[width=0.3\textwidth]{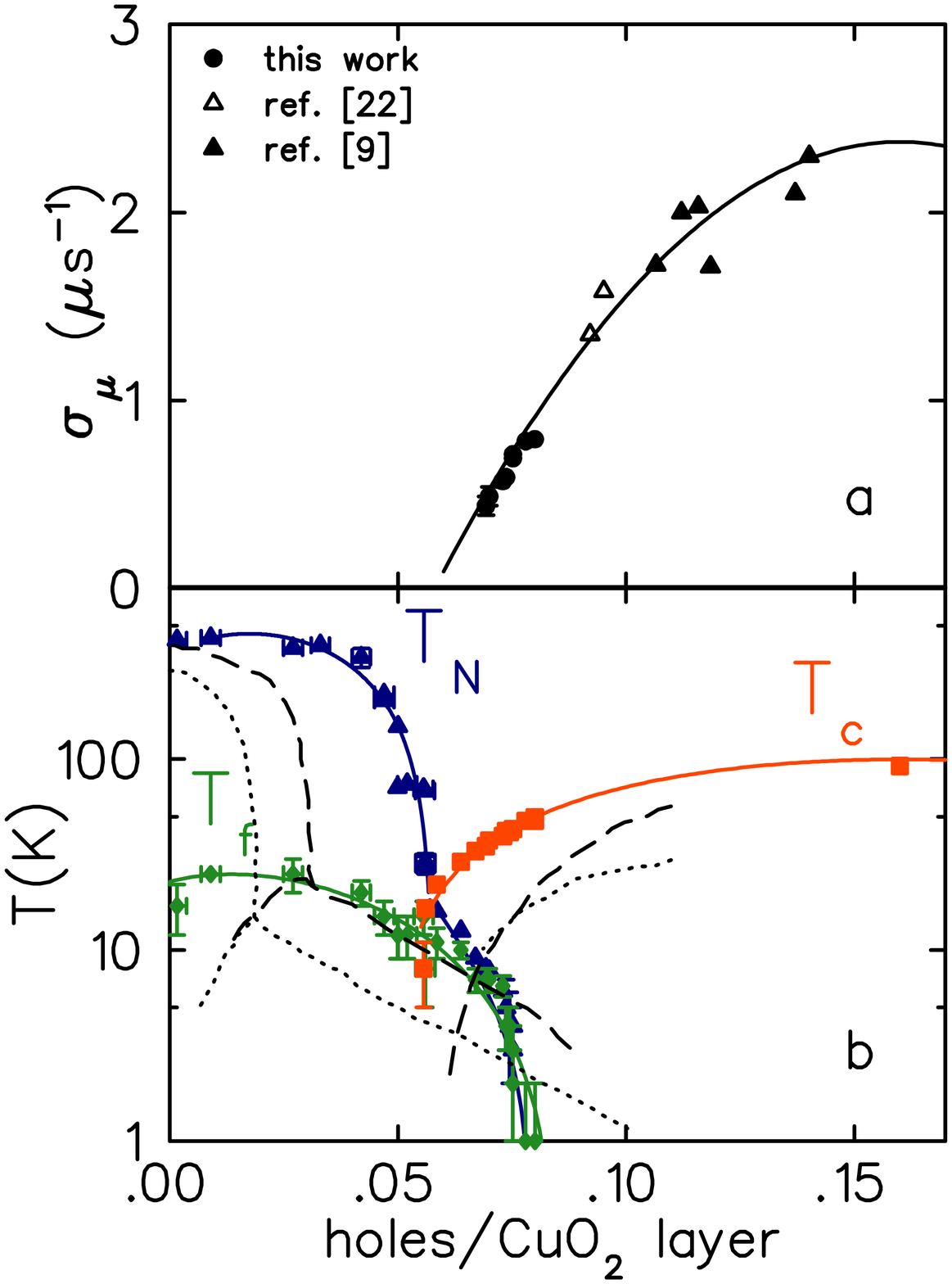}%
 \caption{YBa$_2$Cu$_3$O$_{6+x}$: a) muon Gaussian width $\sigma_\mu$ extrapolated to $T=0$ vs.$h$; b) phase diagram. Solid lines are guides to the eye, dashed (\ybcoca) and dotted (\lsco) lines from ref.~\onlinecite{Niedermayer:1998}.}
 \label{fig:4}
 \end{figure}
Figure \ref{fig:4}b shows the full phase diagram for 123, including our older data. From this plot the critical hole density for superconductivity is $h_c\!\!=\!\!0.055(3)$. Note that $T_f(h)$ extrapolates to zero around $h\!\!=\!\!0.082(3)$, a much lower value than those of \lsco\ and \ybcoca, and very far from the optimal SC value $h\!\!=\!\!0.16$. This agrees with the larger distance between active and charge reservoir layers in the case of 123, which reduces the influence of ionic disorder on magnetism and it suggests the lack of a causal link between AF and SC properties.

The appearance of local internal fields at all muon sites implies nm distance between AF stripes, while superconductivity requires a domain size larger than $\xi\!\!\agt\!\! 10$ nm. How can these two geometrical constraints coexist? Only if AF correlations and SC pairing do not interfere. Metallic clusters nucleate in stripe form already at very low doping  \cite{Borsa:1995,Sanna:2003}, in what resembles the negative of Fig. \ref{fig:3}c. AF cluster spin glass order survives above the critical metallic percolation threshold $h_c$, which corresponds to SC onset, probably because disorder frustrated AF correlations are  harmless to SC pairing. 
 
Finally the dependence of the muon width\cite{Barford:1988} $\sigma_\mu(T=0)\!\!\propto\!\!\lambda^{-2}$ upon $h$ is plotted in Fig.~\ref{fig:4}a, together with a selection of results for higher dopings from ref.~\onlinecite{Savici:2002,Miller:2003} and references therein. The penetration depth is $\lambda\!\!\propto\!\! (n_s/m)^{-\frac{1}{2}}$  hence, neglecting the variation of the effective mass $m$, the data show the dependence of the density of supercarriers $n_s$ on $h$. Our data confirm the parabolic decrease of $n_s(h)$ down to very small values of $h-h_c$.  Our data are also in agreement with the linear portion of the Uemura-plot \cite{Uemura:1991}, thanks to the parabolic shape \cite{Tallon:1995} of  $T_c$ vs. $h$.  The largest measured value of $\lambda(h)$ is 410(20) nm for $h\!\!=\!\!0.070(1)$.

In conclusion we have demonstrated that a nanoscopic coexistence of static stripe-like magnetism and superconductivity takes place in 123 as well. We have carefully mapped the phase diagram and complemented earlier measurements of $\lambda(h)$. Our observation of bulk superconductivity in Y$_{15}$ and Y$_{20}$ below the onset of spontaneous AF internal fields throughout most of the sample suggests that AF order and SC show minimal interference. In view of the nanoscopic coexistence of AF and SC domains it seems that the AF strings are almost transparent for superconducting carriers, which in our view reduces the plausibility for any {\em magnetic} model of high temperature superconductivity.

\begin{acknowledgments}
We gratefully acknowledge the support of G.~Del Fiacco in sample preparation and characterization. We thank G. Guidi and M. Ricc\`o  for fruitful discussions.  RDR acknowledges financial support by FIRB project {\em Materiali magnetici innovativi nanoscopici}. 

\end{acknowledgments}

\bibliography{YBCO}

\end{document}